\documentclass[twocolumn]{article}
\usepackage{}
\usepackage{geometry}
\usepackage{bibspacing}
\usepackage{amssymb} %% two column, final layout
\usepackage[implicit=false]{hyperref}
\usepackage{bbm}
\usepackage{mathrsfs}
\usepackage{amsfonts}
\usepackage{amssymb}
\usepackage{graphicx}
\usepackage{cite}
\usepackage[linesnumbered,ruled,lined]{algorithm2e}
\usepackage{array}
\usepackage{subfigure}
\usepackage{multirow}
\usepackage{color}
\usepackage{amsmath}
\usepackage{bm}
\usepackage{upgreek}
\usepackage{float}

\usepackage{cite}
\usepackage{authblk}

\begin{document}

\title{Multispectral imaging using a single bucket detector}

\author[]{Liheng Bian,$^1$ Jinli Suo,$^1$ Guohai Situ,$^2$ Ziwei Li,$^1$ Feng Chen,$^1$ and Qionghai Dai$^1$
%\thanks{qhdai@tsinghua.edu.cn}
}

\affil[]{
$^1$Department of Automation, Tsinghua University, Beijing 100084, China\\
$^2$Shanghai Institute of Optics and Fine Mechanics, Chinese Academy of Sciences, Shanghai 201800, China
}

\maketitle

\begin{abstract}
%Single pixel imaging techniques enable to capture a scene with a single pixel detector. Because single pixel detectors can not resolve light signals of different wavelength, state-of-the-art single techniques using a single pixel detector can only produce gray scale images.
{\bf
Current multispectral imagers suffer from low photon efficiency and limited spectrum range. These limitations are partially due to the technological limitations from array sensors (CCD or CMOS), and also caused by separative measurement of the entries/slices of a spatial-spectral data cube. Besides, they are mostly expensive and bulky.
To address above issues, this paper proposes to image the 3D multispectral data with a single bucket detector in a multiplexing way.
Under the single pixel imaging scheme, we project spatial-spectral modulated illumination onto the target scene to encode the scene's 3D information into a 1D measurement sequence. Conventional spatial modulation is used to resolve the scene's spatial information. To avoid increasing requisite acquisition time for 2D to 3D extension of the latent data, we conduct spectral modulation in a frequency-division multiplexing manner in the speed gap between slow spatial light modulation and fast detector response.
Then the sequential reconstruction falls into a simple Fourier decomposition and standard compressive sensing problem.
A proof-of-concept setup is built to capture the multispectral data (64 pixels $\times$ 64 pixels $\times$ 10 wavelength bands) in the visible wavelength range (450nm--650nm) with acquisition time being 1 minute. The imaging scheme is of high flexibility for different spectrum ranges and resolutions. It holds great potentials for various low light and airborne applications, and can be easily manufactured production-volume portable multispectral imagers.}
%The reported system is simple and cost saving. Due to the utilized single pixel imaging scheme, the system owns high signal to noise ratio, large spectral range,.
%We discussed its widely potential applications, especially in microscopy.
%The reported technique is cost saving for using only one single pixel detector, and is highly efficient by using the redundancy of single pixel detector's high sampling speed.
\end{abstract}

\section{Introduction}

% 从现有光谱仪的缺点引入：价格贵，低snr，速度慢，体积大，光谱范围短等。然后引入使用单像素能解决这些问题。最后可以在discussion里面再讨论可以通过散射介质成像什么的。
%Spectroscopy studies the interaction between matter and light by analyzing the spectrum of reflected or scattered light from the target object \cite{skoog1980principles}. Therefore, spectroscopy is extremely useful in almost every field of science wherever light-mater interaction exists \cite{garini2006spectral, bacon2004miniature}.
%Generally speaking, spectroscopic instruments can be classified into two types: single point spectrometers which measure the spectrum of a single
%scene point, and multispectral imagers that measure the 3D spatial-spectral data cube (2 spatial dimensions and 1 spectral dimension) of a whole scene. Due to the spatial resolving ability, multispectral imagers are preferred for studying properties of non-uniform scenes.
Multispectral imaging is a technique capturing a spatial-spectral data cube of a scene, which contains multiple 2D images under different wavelengths. Possessing both spatial and spectral resolving abilities, multispectral imaging is extremely vital for surveying a scene and extracting detailed information \cite{garini2006spectral}.
Current multispectral imagers mostly utilize dispersive optical devices (e.g., prism and optical grating) or narrow band filters to separate different wavelengths, and then use an array detector to separately measure them \cite{gat2000imaging,james2007spectrograph, bao2015colloidal}. Using the compressive sensing technique, multispectral images can be multiplexed together to reduce the number of shots \cite{arce2014compressive}.
Another kind of multispectral imaging method is Fourier spectroscopy technique\cite{garini1996spectral}. This approach uses an interferometer to divide the incoming beam into two halves with variable optical path difference, and generate varying interference intensity at each spatial point. The spectral information can be extracted by applying Fourier transform to the intensities measured by an array detector.
Despite the diverse principles and setups of the above multispectral imaging instruments, the photons are detected separately either in the spatial or spectral dimension using array detectors. Therefore, these multispectral imagers are photon inefficient and spectrum range limited. Besides, they are usually bulky \cite{bao2015colloidal} and highly expensive (for example, more than $\$$50000 for NIR-SWIR range multispectral imagers \cite{valerie2015hyperspectral}). %due to the utilized fine optics, array detectors and sophisticated systems.
These disadvantages prevent them from wide practical applications.

\begin{figure}[!t]
  \centering
  \includegraphics[width=\linewidth]{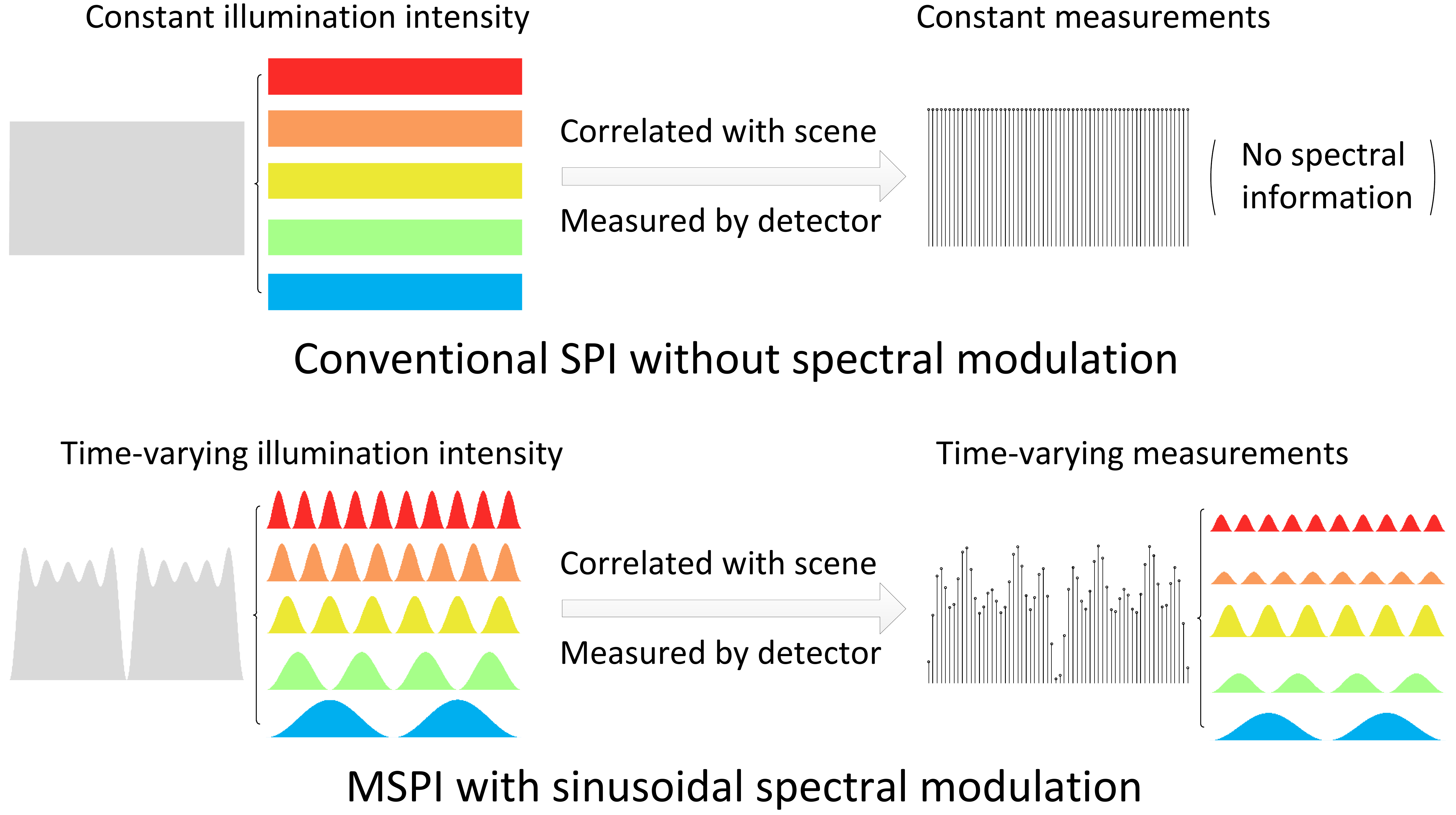}
  \caption{{\footnotesize{\bf Illustration of the difference between conventional SPI and the proposed MSPI}. Due to the response speed gap between a bucket detector (MHz or GHz) and a spatial light modulator (no higher than KHz), the detector can collect a dense sequence of measurements during elapse of each spatially modulated pattern. In conventional SPI, given a spatial pattern, its light intensity and corresponding measurements are constant. Thus no spectral information could be extracted from the sequence. Differently, in MSPI both the intensity and measurements are time-varying, since the intensity of each spectral component changes sinusoidally with their own frequencies over time. The speed gap enables us to multiplex and demultiplex scene's spectral components from the measurement sequence during elapse of each spatial pattern.}}
  \label{fig:Illustration}
\end{figure}

\begin{figure*}[!t]
  \centering
  \includegraphics[width=0.9\linewidth]{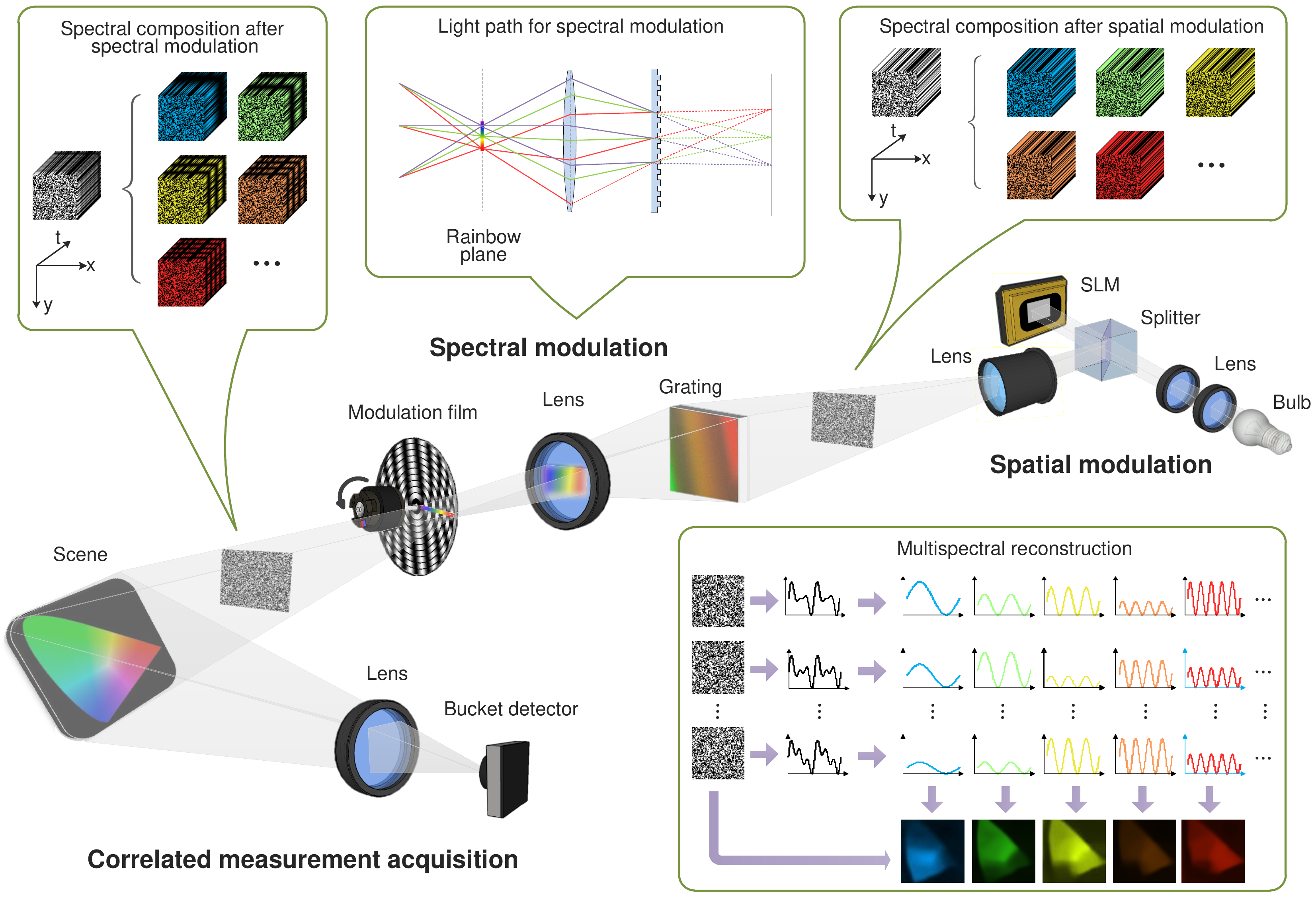}
  \caption{{\footnotesize{\bf Schematic of the proposed single-pixel multispectral imaging system}. The broadband light from the high power bulb is spatially modulated by a spatial light modulator (SLM) to generate a series of 2D random patterns. Next, the spectra of the 2D patterns are distributed into a rainbow stripe, and modulated by a rotating film before transformed back to 2D patterns. After both the spatial and spectral modulations, the incident illumination is tailored structurally in three dimensions---random in the 2D spatial dimensions and sinusoidal along the spectral dimension. Then the patterns illuminate the target scene to encode both its spatial and spectral information. Finally a bucket detector is utilized to measure the correlated signals. In the sequential reconstruction process, the spectral response signals are decoded by Fourier decomposition, while the spatial information are demodulated by a compressive sensing based reconstruction algorithm. Details of the modulations and demodulations are shown in the insets.}}
  \label{fig:System}
\end{figure*}

Differently, single pixel imaging (SPI)\cite{duarte2008single, shapiro2008computational} provides a promising scheme being able to address the above issues of current multispectral imaging instruments.
Using a bucket detector instead of expensive and bulky CCD or CMOS, SPI systems are of low cost, compact, and own wider spectral detection range \cite{edgar2015simultaneous}. Besides, SPI collects all the lights interacted with the scene to a single detection unit. Thus it is more photon efficient \cite{multiplexing, davis2011multivariate, morris2015imaging}. What's more, SPI is flexible, meaning that it attaches no requirement on the light path between scene and the detector, providing that all the interacted lights are collected to the detector \cite{Nature_2015}.
In the past years, SPI has achieved great success in 2D imaging and various applications \cite{sun20133d, tian2011fluorescence, clemente2010optical, zhao2012ghost, cheng2009ghost, magana2013compressive}.

To produce advantages of the SPI scheme in multispectral imaging, there are two intuitional ways. One is to resolve the spectra of the collected measurements at the detector. Existing such methods include i) directly replacing the bucket detector with a spectrometer \cite{His_Spectrometer, His_OpticalSpectrumAnalyzer}, and ii) using light filters \cite{His_Filter_1, His_Filter_2} or dispersive optical devices \cite{august2013compressive, His_DichroicPrism} to separate signals of different wavelengths, and then measure them separately. Another straightforward way is to directly extend the 2D spatial modulation to 3D spatial-spectral modulation using two spatial light modulators. However, this would largely increase requisite projections \cite{multiplexing} and corresponding computation complexity for reconstruction.
In a word, since a single bucket detector cannot distinguish different spectra, the above methods needs either high commercial cost or geometrically increasing projections and computational cost.

In this paper, we propose a novel single pixel multispectral imaging technique, termed as multispectral single pixel imaging (MSPI), without increasing requisite projections and capturing time compared to conventional SPI. The main difference between conventional SPI and MSPI is illustrated in Fig. \ref{fig:Illustration}.
Utilizing the fact that the response speed of a bucket detector (MHz) is magnitudes faster than illumination patterning (KHz)\cite{sun20133d, His_Filter_1, suo2015self}, we encode the spectral information into this speed gap.
Specifically, the proposed MSPI technique introduces spectrum-dependent sinusoidal intensity modulation to the lights, during the elapse of each spatially modulated pattern.
Thus, different spectrum bands are multiplexed together into the 1D dense measurements at the bucket detector in a frequency-division multiplexing manner.
Since the response signals of different bands displays distinct dominant frequencies in the Fourier domain, we conduct a simple Fourier decomposition to separate multispectral response signals from each other. Last, the compressive sensing algorithm \cite{duarte2008single} is applied to these signals in different wavelength bands to reconstruct the latent multispectral data.
%The multiplexing of both spatial and spectral information makes MSPI achieve high light efficiency, and thus own great potential applications in low light conditions such as biological imaging \cite{morris2015imaging} and Raman imaging \cite{davis2011multivariate}.
The spectral multiplexing and demultiplexing based on Fourier decomposition can suppress system noise effectively, and thus produces high robustness to noise and ensures high reconstruction quality.

\begin{figure*}[!t]
  \centering
  \includegraphics[width=0.9\linewidth]{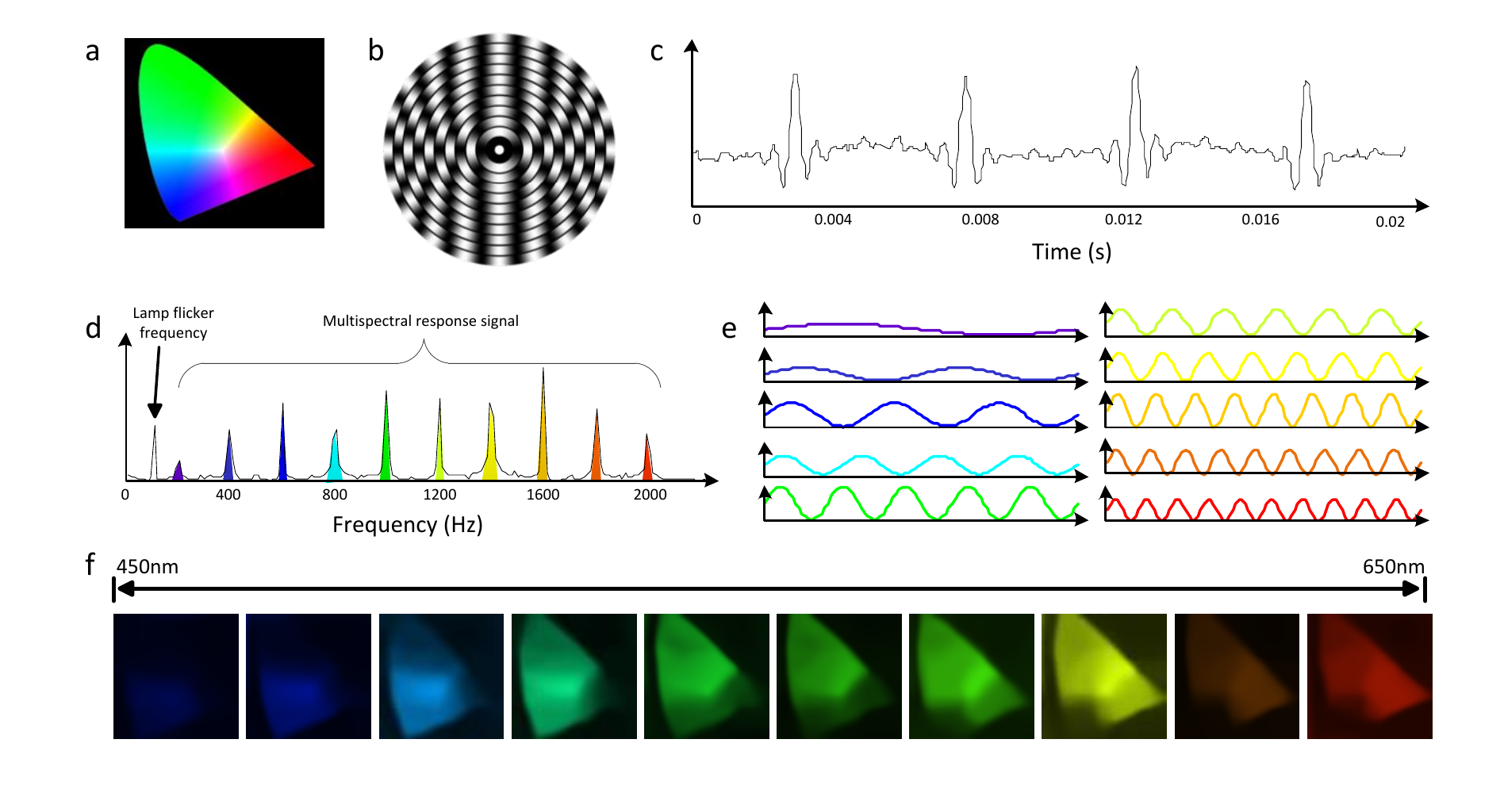}
  \caption{{\footnotesize{\bf Multispectral imaging results on a color scene.} (a) is the target color scene (a printed film of CIE 1931 color space). (b) is the sinusoidal modulation film used in our setup. While the rainbow spectrum is converged along the radius of the film, different wavelengths are modulated with different sinusoidal periods as the film rotates. (c) shows exemplar recorded correlated measurements corresponding to a specific projecting pattern. (d) is the Fourier decomposition of the measurements, which displays several dominant frequencies. The coefficients of the dominant frequencies correspond to the response signals' strengths of specific wavelengths. (e) shows the decomposed sequences for different spectrum bands, while (f) presents the final reconstructed 2D multispectral images ($64\times64$ pixels) corresponding to 10 narrow bands.}}
  \label{fig:Results_Charactor}
\end{figure*}

MSPI owns a lot of potential applications in various fields of science. Due to its high photon efficiency and robustness to noise, MSPI could be used in low light conditions, such as fluorescence microscopy \cite{studer2012compressive} and Raman imaging \cite{davis2011multivariate}. Besides, the utilized SPI scheme enables MSPI system to be of compact size and low weight. This is beneficial for a lot of airborne applications, including geologic mapping, mineral exploration, agricultural assessment, environmental monitoring, and so on \cite{shaw2003spectral}. Moreover, MSPI applies to a large spectral range and is of low cost, thus can be used for production-volume portable devices for daily use.

\section{Results}

\paragraph{Experimental setup.} MSPI builds on the SPI scheme. In SPI, the incident uniform illumination is patterned by a spatial light modulator (SLM), and then projected onto the target scene to multiplex its spatial information. Simultaneously, a bucket detector is used to collect the encoded measurements. Afterwards, the compressive sensing algorithm \cite{duarte2008single} retrieves the spatial information of the target scene computationally.
Under a similar architecture, MSPI adds an extra spectral modulation to the incident light to resolve the scene's spectral resolving information. The principle of the proposed MSPI system is sketched in Fig. \ref{fig:System}. On a whole, MSPI projects spatial-spectral modulated light beam to modulate corresponding information of the target scene, and collects the correlated lights with a single bucket detector. Integrating both spatial and spectral modulation, MSPI could resolve a spatial-spectral 3D data cube of the target scene computationally, as displayed in the bottom right inset of Fig. \ref{fig:System}.

We built a proof-of-concept setup to verify the functionality of MSPI, as shown in Fig. \ref{fig:System}. A broadband light source (Epson white 230W UHE lamp) is converged and collimated via a set of optical elements for succeeding modulation. For spatial modulation, we use a digital micromirror device (DMD, Texas Instrument DLP Discovery 4100, .7XGA), which can switch binary patterns at a given frequency (20kHz maximum) with clean-cut pattern transition. The intensity of the spatial illumination pattern is temporally constant for now, as visualized in the top right inset. The illumination pattern is then diverged by a projector lens (Epson, NA 0.27) for successive spectral modulation.
The spectral modulation module is similar to the agile multispectral optical setup \cite{Agile_1}, with the light path displayed in the top middle inset.  %consists of an optical grating, a convex lens and a modulation film.
Specifically, an optical grating (600 grooves, $\phi=50mm$) is placed on the focal plane of the spatial illumination patterns. Then a convex lens collects the first order dispersed spectrum, and focuses it onto the rainbow plane, where a round film printed with sinusoidal annuluses owning different periods spectral modulation is placed for spectral modulation. The rainbow spectrum stretches along the film's radius. Driven by an electric motor rotating at a constant speed (around 6000 r/min), the film realizes a wavelength-dependent intensity modulation to the spectra, i.e., different wavelengths own different temporally sinusoidal intensity variations, as visualized in the top left inset of Fig.~\ref{fig:System}. After both spatial and spectral modulation, the illumination patterns interact with the scene, and we use a bucket detector (Thorlabs PDA100A-EC Silicon photodiode, 340-1100$nm$) together with a 14-bit acquisition board ART PCI8514 to capture the correlated lights.
For reconstruction, we first conduct spectral demultiplexing using simple fast Fourier transform (computation complexity is $\mathcal{O}(n \log n)$), and then reconstruct multispectral scene images using the linearized alternating direction method \cite{ADM} (computation complexity is $\mathcal{O}(n^3)$) to solve the compressive sensing model. Readers are referred to the Methods section for reconstruction details.

In the following experiments, 3000 spatially random modulated patterns (each owning 64$\times$64 pixels) are sequentially projected onto the target scene. The frame rate of the DMD is set to be 50Hz, and the sampling rate of the bucket detector is 100kHz. We utilize the novel self-synchronization technique in \cite{suo2015self} to synchronize the DMD and the detector. It takes us around 1 minute for data acquisition.

\begin{figure*}[!t]
  \centering
  \includegraphics[width=0.9\linewidth]{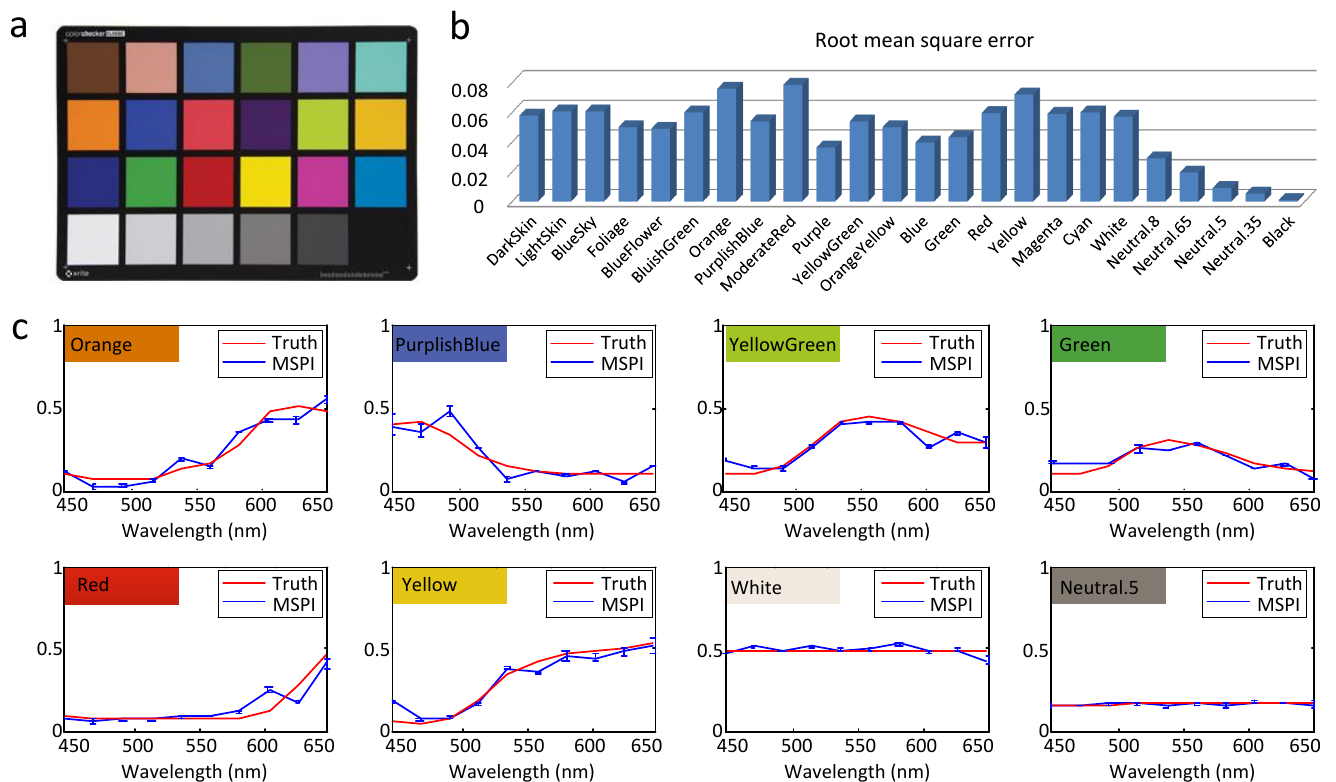}
  \caption{{\footnotesize{\bf Quantitative analysis on the imaging accuracy of MSPI.} (a) is the target scene---X-Rite standard color checker, which consists of 24 swatches owning different known spectra. We use MSPI to image the color checker and obtain 10 multispectral images (450nm-650nm), and calculate the recovered spectrum of each swatch as the average of all corresponding pixels' spectra. (b) presents the reconstruction error of the swatches in terms of root mean square error. (c) shows direct comparison between recovered spectra and their ground truth counterparts on several representative swatches. The standard deviation of each band is also calculated and shown as blue bars. Both the small reconstruction error and deviation validate the accuracy and robustness of MSPI.}}
  \label{fig:Results_ColorBar}
\end{figure*}

\paragraph{Multispectral imaging results of MSPI.} We first apply the proposed MSPI technique to capture the multispectral images of a scene with rich color. Here we use a printed 'CIE 1931 color space' image with wide spectrum range (see Fig.~\ref{fig:Results_Charactor}(a)) to demonstrate the effectiveness of the proposed approach. In this experiment, the rainbow spectrum ranges from 450nm to 650nm. The length of the rainbow stripe is around 23mm, and we discretize it into 10 narrow bands, by printing 10 2mm annular rings with sinusoidal periods varying from 2 to 20 (as shown in Fig. \ref{fig:Results_Charactor}(b)).
%The selection of the sinusoidal period is theoretically arbitrary, but in implementation we should not using neither too short periods for robustness to the high frequency sensor noise nor too long periods for stable decomposition.
The size of the projected pattern on the film is around 45mm$\times$45mm.

Given an exemplar spatial pattern, the recorded correlated measurements from the single pixel detector are plotted in Fig.~\ref{fig:Results_Charactor}(c), and its corresponding Fourier coefficients are displayed in Fig.~\ref{fig:Results_Charactor}(d). One can see that there exist several dominant peaks, which comes from the sinusoidal codes of corresponding frequencies (the 60Hz peak comes from the lamp flicker due to voltage fluctuations). The magnitudes of these peaks are exactly the strengths of the response signals of corresponding spectrum bands. The other small fluctuations of the Fourier coefficients are caused by system noise. From this we can see that although the multispectral response signals are corrupted with system noise in the temporal domain, they are clearly distinguishable in Fourier space.
Therefore, we can easily demultiplex multispectral response signals from each other and suppress system noise by a simple Fourier decomposition (see the Methods section for more details), and the results are shown in Fig.~\ref{fig:Results_Charactor}(e). The frequencies match exactly with the multiplexing codes printed on the film.
After response signal demultiplexing, we can recover the single-band images separatively using the compressive sensing based algorithm. The reconstructed 10 multispectral scene images are shown in Fig.~\ref{fig:Results_Charactor}(f), we integrate which with the Canon EOS 5D MarkII camera's RGB response curves \cite{jiang2013space} for better visualization. The pleasant results verify the effectiveness of the proposed MSPI.

\paragraph{Analysis on the performance and robustness of MSPI.}
To quantitatively demonstrate the performance of MSPI, we acquire the multispectral data of a X-Rite standard color checker (see Fig.~\ref{fig:Results_ColorBar}(a)) using MSPI, and conduct quantitative analysis on the reconstruction accuracy. In implementation, we introduce a pair of cylinder mirrors to match the shape of the light beam with that of the color checker (125mm $\times$ 90mm). For each swatch on the checker, we average all the pixels' reconstructed spectra as the swatch's reconstruction spectrum. Reconstruction error in terms of root mean square error among the 10 spectral bands is calculated for each swatch, and the results of all the 24 swatches are shown in Fig.~\ref{fig:Results_ColorBar}(b). For more direct comparison, we show the spectrum comparison between the reconstruction and the ground truth of several representative swatches in Fig.~\ref{fig:Results_ColorBar}(c).
From the small deviation compared to the ground truth, especially the ones with large estimation error (e.g., 'Orange' and 'Yellow'), we can see that the reconstructed spectra of the swatches are compliant with the ground truth.
This experiment largely validates the multispectral reconstruction accuracy of MSPI.
The accuracy benefits from the high precision of spectral demultiplexing (clear-cut discrimination between the Fourier coefficients of signals and noise), as well as the optimization reconstruction algorithm.

%Further, considering the system noise is a factor may affecting the final performance, we specially do an experiment to show the robustness of our system. Generally, there will exist high frequency sensor noise at the photodiode, and the noise can be non-white. From the reconstruction detailed in the XXX Section, we can clearly see that our reconstruction uses only the coefficients at given dominant low frequencies. Differently, the coefficients of sensor noise will concentrate on high frequencies and the zero frequency (for non-white noise). Therefore, the sensor noise can be canceled out simply by our Fourier domain regularization. To validate this conclusion, we vary the illumination strength to change the signal-noise-ratio at the single pixel detector, and compare the reconstruction qualities of the multispectral data. Here we decreases the strength by factor of 2 for four times and the corresponding reconstruction (after integrating with the RGB response curves of XXX camera) are shown in Fig.~\ref{fig:Results_ColorBar}(c). One can see that the quality only decreases slightly as the noise level increases. The high robustness to sensor noise is mainly attributed to the clear-cut discrimination between the Fourier coefficients distributions of signal and noise.

\section{Discussion}

This paper proposes a new multispectral imaging technique using a single bucket detector, termed as MSPI. Making use of the speed gap between the slow spatial illumination patterning and the fast detector response, MSPI extends conventional 2D spatial coding to 3D spatial-spectral coding via temporal sinusoidal spectral modulation within each spatial pattern elapse. This technique successfully resolves multispectral information without introducing additional acquisition time and computational complexity to conventional 2D SPI.
%Utilizing Fourier decomposition, we successfully demultiplex different response signals of different wavelengths for subsequent scene reconstruction.
The proposed MSPI holds great potential for developing cheap, compact and high photon efficient multispectral cameras.

The specifications of the spectral modulator are flexible and can be easily customized. First, the width of the printed annuluses on the film determines the spectral resolution and can be adjusted for specific resolutions. Second, we can also use a grating with denser grooves to lengthen the rainbow stripe and raise the spectral resolution alternatively. Third, the multiplexing mode can change easily by designing other film graphs. The sinusoidal spectral modulation utilized in current MSPI system is adopted due to its simplicity. We refer readers to \cite{multiplexing} for more multiplexing methods.
%, and replacing the printed film with programmable spatial light modulators permits more freedom for spectral coding \cite{lin2014dual}

Recalling that the proposed technique is a general scheme for multispectral imaging, it can be conveniently coupled with a variety of imaging modalities (no matter macroscopy or microscopy), by using corresponding optical elements. The scheme is wavelength independent, and users can apply the scheme to other spectrum ranges readily. This is especially important for the wavelengths under which array sensors are costly or unavailable. In addition, similar to the system in \cite{His_Filter_2}, the modulation can be conducted after the light beam interacted with the target scene. This enables us to analyze the scene's spatial-spectral information without active illumination. One can refer to the supplementary material for details of MSPI under passive illumination, which is of wider applicability.

Although MSPI owns many advantages over conventional multispectral imaging techniques, these benefits come at the expense of a large number of projections and algorithmic reconstruction. In other words, MSPI makes a trade-off of temporal resolution for spatial and spectral resolution. Fortunately, the imaging speed of MSPI can be accelerated utilizing advanced techniques. In terms of data acquisition, current efficiency is mainly limited by the spectral modulator, and we can use a faster rotation motor or denser sinusoidal patterns for acceleration. In terms of reconstruction, considering there exists abundant redundancy among different color channels\cite{park2004color, han2014fast}, we can utilize this cross channel prior in the reconstruction to reduce the requisite projections and thus accelerate imaging speed. The reconstruction time can also be shortened further, because different spectrum bands are reconstructed separately, and we can utilize graphics processing unit (GPU) to reconstruct different channels in a parallel manner.

Besides, current spatial resolution is apparently insufficient for practical applications. Targeting for proof-of-concept and without loss of generalization ability, here we project randomly spatial modulated patterns in the capturing stage, similar to most SPI systems. However, recent studies \cite{Nature_2015, bian2015fourier} show that projecting structural and adaptive patterns instead of random ones can largely improve the spatial resolution while decreasing projections and lowering computation cost. Hence, we can easily improve the spatial resolution under exactly the same scheme.

%Besides, the utilized SPI scheme attaches no requirement on the light path between the scene and the detector, providing that all the photons are collected to the detector. This property brings more tolerance to the light transmission medium than conventional cameras and opens up possibilities for hyperspectral imaging through scattering medium \cite{cheng2009ghost, zhang2010correlated}.

\section{Methods}
The reconstruction of the proposed MSPI technique consists of two main steps, namely spectral demultiplexing and multispectral reconstruction.

\paragraph{Spectral demultiplexing.}
%Making use of the velocity gap between the illumination patterning and detector's response, for each illumination pattern, we can obtain a sampling sequence from the single pixel detector.
Due to the sinusoidal spectral modulation, for a spatially modulated pattern, its measurement sequence from the bucket detector consists of several response signals of different spectra. These response signals own different frequencies of sinusoidal intensity variations. Thus in the Fourier domain, the measurement sequence is composed of several corresponding dominant frequencies. Besides, there exists system noise in the measurements, we assume which to be stochastic and zero-mean. In the Fourier domain, the noise mainly locates at high frequencies. Adopting simple Fourier decomposition \cite{bloomfield2004fourier}, we could separate the response signals from each other and from the measurement noise.

Mathematically, the Fourier decomposition describes a time series as a weighted summation of sinusoidal functions at different frequencies. A captured measurement sequence $\{y_0, \cdots, y_{T-1}\}$ (captured with a given spatial illumination pattern) can be represented by a series of sinusoidal functions as
\begin{eqnarray}\label{eqs:Fourier_1}
y_t = b_0 + \sum_{i=1}^{T/2}\left\{b_i\sin(\frac{2\pi i}{T}t + \phi_i )\right\}.
\end{eqnarray}
In this equation, $b_0 = \frac{1}{T}\sum_{t=0}^{T-1}y_t$, $b_i = \frac{2}{T}\sqrt{\left[\sum_{t=0}^{T-1}y_t\cos(\frac{2\pi i}{T}t)\right]^2+\left[\sum_{t=0}^{T-1}y_t\sin(\frac{2\pi i}{T}t)\right]^2}(i>0$), and $\phi_i = \arctan \frac{\sum_{t=0}^{T-1}y_t\sin(\frac{2\pi i}{T}t)}{\sum_{t=0}^{T-1}y_t\cos(\frac{2\pi i}{T}t)}$.
Specifically, $b_0$ is the direct current component indicating the average of the measurements, while $b_i (i>0)$ indicates the energy of the $i$th sinusoidal function with modulation frequency $\frac{i}{T}$.
As stated before, each spectrum band corresponds to one specific sinusoidal modulation frequency. Thus, the above coefficients at the specific frequencies are exactly the response signals corresponding to the spectral bands.
Here we adopt fast Fourier transform (FFT) to transfer the measurements into Fourier domain, with computation complexity being $\mathcal{O}(n \log n)$. Then we demultiplex the response signals corresponding to different spectrum bands by finding the local maximum coefficients around corresponding Fourier frequencies.

By doing FFT to each measurement sequence, we obtain a set of response signals for each spectral band.
Mathematically, assuming that the wavelength $\lambda$ is modulated with sinusoidal frequency being $\frac{j}{T}$, we can obtain a response signal $b_j$ from the measurement sequence corresponding to one projecting pattern. Considering that we project $m$ patterns, we can get $m$ response signals of the wavelength $\lambda$. In the following, we indicate the response signal set as a row vector ${\bf b}_\lambda\in \mathbb{R}^{m}$. Each entry in ${\bf b}_\lambda$ corresponds to a response signal of the band $\lambda$ for one pattern.

%\begin{eqnarray}\label{eqs:Fourier_1}
%y_t &=& \sum_{i=0}^{T/2}\left\{\alpha_i\cos(\frac{2\pi i}{T}t)+\beta_i\sin(\frac{2\pi i}{T}t)\right\} \\ \nonumber
%&=& b_0 + \sum_{i=1}^{T/2}\left\{b_i\sin(\frac{2\pi i}{T}t + \phi_i )\right\} \nonumber
%\end{eqnarray}
%where $\alpha_i = \frac{2}{T}\sum_{t=0}^{T-1}y_t\cos(\frac{2\pi i}{T}t)$, $\beta_i = \frac{2}{T}\sum_{t=0}^{T-1}y_t\sin(\frac{2\pi i}{T}t)$.

\paragraph{Multispectral reconstruction.} After demultiplexing response signals of different wavelengths, the reconstruction is implemented separately for each wavelength band. For band $\lambda$, we assume the spatial pixel number of each illumination pattern is $n$, and denote the pattern set as ${\bf A}\in \mathbb{R}^{m\times n}$ (each pattern is represented as a row vector).
The multispectral scene images own the same spatial resolution as the illumination patterns, and is denoted as ${\bf x}_\lambda\in \mathbb{R}^{n}$ for the wavelength $\lambda$. %Thus, the signal formation can be represented as $Ax = b$. This is the constraint of the reconstruction model.

To reduce the number of requisite projections, we choose to conduct reconstruction under the framework of compressive sensing\cite{duarte2008single}. The reconstruction is performed by solving the following optimization problem:
\begin{eqnarray}\label{eqs:Model}
\{{\bf x}_\lambda^*\} = \arg\min && ||\psi({\bf x}_\lambda)||_1 \\ \nonumber
s.t. && {\bf Ax}_\lambda = {\bf b}_\lambda.\nonumber
\end{eqnarray}
The definition of the objective comes from a sparsity prior: natural scene images are statistically sparse when represented with an appropriate basis set (e.g. the discrete cosine transform basis) \cite{Prior}. %This means that a scene image contains many coefficients close or equal to zero when represented in the basis.
We use $\psi({\bf x}_\lambda)$ to denote the coefficient vector, with $\psi$ being the mapping operator to the transformed domain, and minimize its $l_1$ norm to force the sparsity. Eq.~\ref{eqs:Model} is a standard $l_1$ optimization problem, and there exist many effective algorithms to solve it. Here we use the linearized alternating direction method \cite{ADM} to obtain the optimal ${\bf x}_\lambda^*$, with computation complexity being $\mathcal{O}(n^3)$. This results in the final reconstructed scene image corresponding to the specific wavelength band $\lambda$. After doing the above reconstruction to all the wavelength bands, we get multispectral images of the target scene.

\bibliographystyle{unsrt}
\footnotesize
\bibliography{MultispectralCGI}% Produces the bibliography via BibTeX.

\vspace{8mm}
\noindent {\bf \Large Acknowledgements}
\vspace{3mm}

We thank Yuwang Wang, Ziyan Wang and Jing Pu for their valuable discussions and help. This work was supported by the National Natural Science Foundation of China (Nos. 61171119, 61120106003, and 61327902).

%\vspace{8mm}
%\noindent {\bf \Large Author contributions}
%\vspace{3mm}
%
%Liheng Bian and Jinli Suo proposed the idea and designed the experiments. Liheng Bian and Ziwei Li built the setup and conducted the experiments. All the authors contribute to writing and revising the manuscript, and convolved in the discussions during the project.
%
%\vspace{8mm}
%\noindent {\bf \Large Additional information}
%\vspace{3mm}
%
%{\bf Competing financial interests:} The authors declare no competing financial interests.

%\vspace{3mm}
%{\bf Reprints and permission} information is available online at http://npg.nature.com/
%reprintsandpermissions/
%
%\vspace{3mm}
%{\bf How to cite this article:} Bian, L. et al. xxx xxx xxx xxx xxx xxx xxx xxx.
%Nat. Commun. x:xxxx doi: xx.xxxx/ncommsxxxx (20xx).

\end{document}